\documentclass[aps,prb,twocolumn,showpacs,groupedaddress]{revtex4-1}
\usepackage{graphics}
\usepackage{graphicx}
\usepackage{dcolumn}
\usepackage{bm}
\usepackage{float}
\usepackage{subfigure}

\begin{document}

\title{Unusual Phase Transitions and Magnetoelastic Coupling in TlFe$_{1.6}$Se$_2$ Single Crystals}

\author{Brian C. Sales$^1$, Michael A. McGuire$^1$, Andrew F. May$^1$, Huibo Cao$^2$, Bryan C. Chakoumakos$^2$ and Athena S. Sefat$^1$}
\affiliation{Materials Science and Technology Division, Oak Ridge National Laboratory, Oak Ridge, TN 37831}
\affiliation{Neutron Scattering Science Division, Oak Ridge National Laboratory, Oak Ridge, TN 37831 USA}
\date{\today}

\begin{abstract}
Structural, magnetic, electrical transport, and heat capacity data are reported for single crystals of TlFe$_{1.6}$Se$_2$. This compound crystallizes in a tetragonal structure similar to the ThCr$_2$Si$_2$ structure, but with vacancies in the Fe layer. The vacancies can be ordered or disordered depending on temperature and thermal history. If the vacancies are ordered, the basal plane lattice constant increases from \textit{\textbf{a}} to $\sqrt{5}$\textit{\textbf{a}}.  Antiferromagnetic order with the Fe spins along the \textit{\textbf{c}} axis occurs below $T_N \approx$ 430\,K as shown by single crystal neutron diffraction and the magnetic structure is reported.  In addition, for the vacancy ordered crystal, two other phase transitions are found at $T_1 \approx$ 140\,K, and $T_2 \approx$ 100\,K. The phase transitions at $T_1$ and $T_2$ are evident in heat capacity, magnetic susceptibility, resistivity data, \textit{\textbf{a}} and \textit{\textbf{c}} lattice parameters, and in the unusual temperature dependence of the magnetic order parameter determined from neutron scattering. The phase transitions at $T_1$ and $T_2$ result in significant changes in the magnetic moment per iron, with 1.72(6)\,$\mu_B$ observed at 300\,K, 2.07(9)\,$\mu_B$ at 140\,K, 1.90(9)\,$\mu_B$ at 115\,K, and 1.31(8)\,$\mu_B$ for 5\,K if the same ``block checkerboard'' magnetic structure is used at all temperatures. The phase transitions appear to be driven by small changes in the \textit{\textbf{c}} lattice constant, large magnetoelastic coupling, and the localization of carriers with decreasing temperature.
\end{abstract}


\maketitle

\section{Introduction}

\begin{figure}
	\centering
\includegraphics[width=3in]{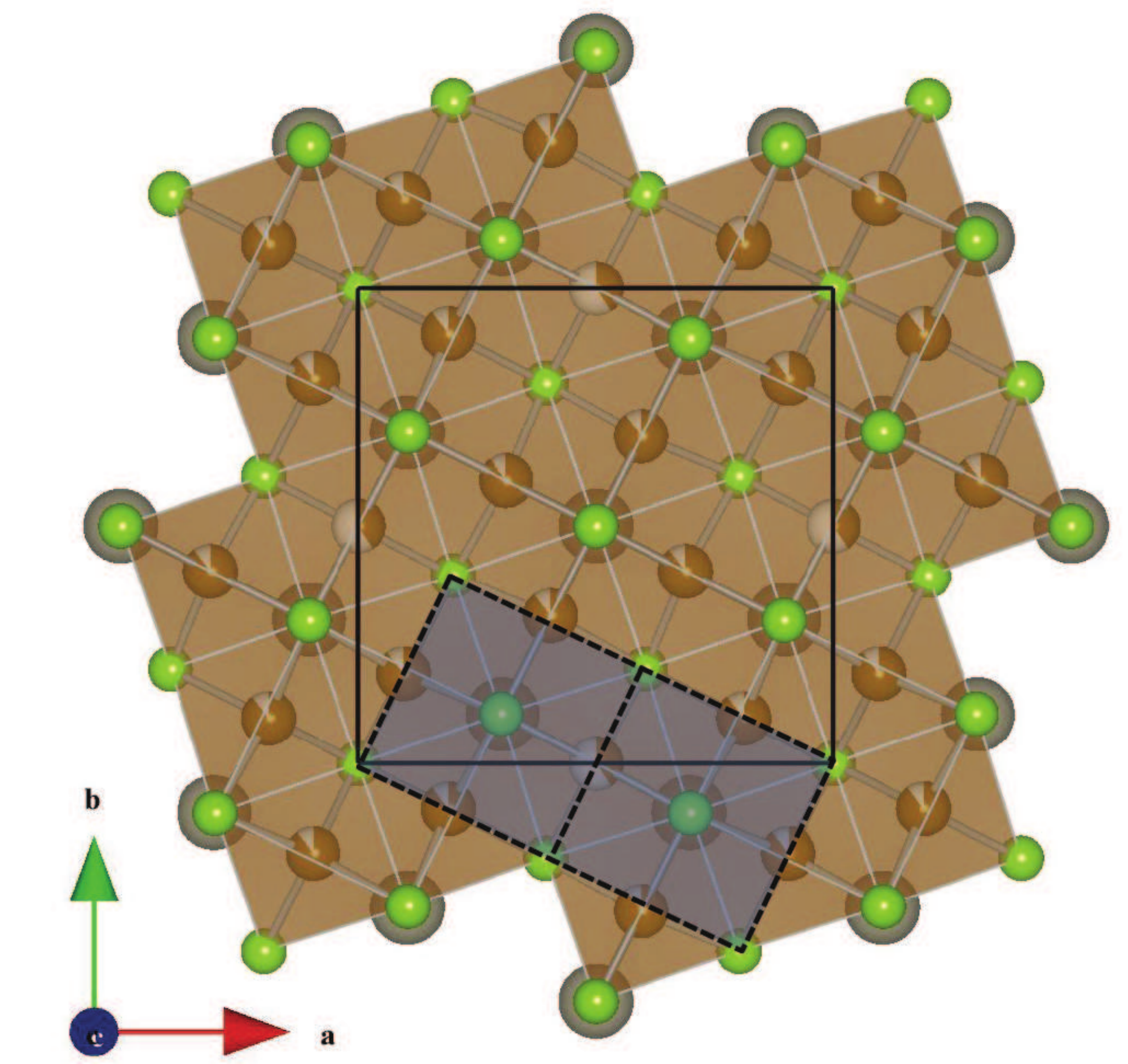}
\caption{(color online) Projection of the crystal structure of TlFe$_{1.6}$Se$_2$ along the \textit{\textbf{c}} axis; spheres represent Tl (gray), Se (green), and Fe (brown) where the Fe site occupancies are shown by partial coloring. The basic tetragonal ThCr$_2$Si$_2$ cell (dashed) is shown as well as the $\sqrt{5}$\textit{\textbf{a}} $\times$ $\sqrt{5}$\textit{\textbf{a}} superscell (solid) that results from vacancy ordering.}
	\label{fig:1}
\end{figure}

One of the early puzzles in the initial electronic structure calculations of the iron-arsenide based superconductors was the difficulty in reconciling the magnetic moment calculated using density functional theory (DFT) and the experimental Fe-As distance. From these early calculations it was evident that the Fe magnetic moment was extremely sensitive to this distance, suggesting remarkably strong magnetoelastic coupling in this family of materials \cite{Singh2009418,Johnston2010Review}. Theoretically, it was shown that the hypothetical compound TlFe$_2$Se$_2$ should also exhibit strong magnetoelastic coupling,\cite{ZhangSingh2009} although experimentally the Fe layer is never completely filled. The compound TlFe$_{1.6}$Se$_2$ crystallizes in a variant of the ThCr$_2$Si$_2$ structure type,\cite{Haggstrom1986} very similar to the structure of the heavily studied ``122'' superconductors such as Ba(Fe$_{0.92}$Co$_{0.08}$)$_2$As$_2$.\cite{Johnston2010Review,SafetPRL2008} It has been shown more recently that in spite of some vacancies in the Fe layer, replacing part or all of the Tl with K, Rb, or Cs results in bulk superconductivity \cite{GuoPRB2010,Wang1012.5525,Maziopa1012.3637,Fang1012.5236} with a Tc in excess of 30\,K. The vacancies in the Fe layer can order, as was first discussed by Sabrowsky \cite{Sabrowsky1986} and H\"{a}ggstr\"{o}m.\cite{Haggstrom1986}  H\"{a}ggstr\"{o}m et al.\cite{Haggstrom1986} found that in TlFe$_{1.6}$Se$_2$, vacancies in the Fe layer order forming a $\sqrt{5}a$ superstructure with the supercell (') related to the subcell by \textit{\textbf{a}}' = 2\textit{\textbf{a}} + \textit{\textbf{b}}, \textit{\textbf{b}}' = -\textit{\textbf{a}} + 2\textit{\textbf{b}}, \textit{\textbf{c}}' = \textit{\textbf{c}}, based on the tetragonal ThCr$_2$Si$_2$ subcell with lattice constant \textit{\textbf{a}} and \textit{\textbf{c}} (see Fig. \ref{fig:1}).

In the present work, the crystal and magnetic structure, electrical transport, heat capacity and magnetism of TlFe$_{1.6}$Se$_2$ single crystals are investigated as a function of temperature. Bulk \textit{\textbf{c}}--axis antiferromagnetism is found below $T_N$$\approx$ 430\,K, in agreement with single crystal Mossbauer data.\cite{Haggstrom1986} In addition, two other phase transitions are found at $T_1$ $\approx$ 140\,K, and $T_2$ $\approx$ 100\,K.  From room temperature to 150\,K the resistivity of TlFe$_{1.6}$Se$_2$ is between 0.006--0.010 $\Omega$ cm, and the magnetic moment per Fe peaks at 2.07(9)$\mu_B$ at 140\,K (based on single crystal neutron diffraction). However, below 100\,K, the resistivity begins to rapidly increase and the apparent moment per Fe is suppressed at the lowest temperature to a value of 1.31(8)$\mu_B$.

\section{Synthesis}

Single crystals of TlFe$_{1.6}$Se$_2$ were grown from a melt using a modified Bridgman technique. Millimeter sized thallium pieces (99.99\%), selenium shot (99.9999\%), and iron pieces (99.99\%) were loaded in a helium glove box into a carbonized Bridgman silica tube (1.5\,cm inner diameter by 12\,cm long, total mass of elements $\approx$ 20\,g) and then sealed under vacuum. The nominal starting composition was Tl:2Fe:2Se. The carbonized tube was then sealed inside a second silica tube in case the inner tube cracked on cooling. The mixture was heated in a box furnace to 1070\,$^{\circ}$C at 2$^{\circ}$C/min, held for 24\,h and cooled at 6$^{\circ}$C/h down to near 400$^{\circ}$C and held for 24\,h, and then finally furnace cooled to room temperature.  The silica ampoule was placed in the furnace to use the small natural temperature gradient of the box furnace. Careful single crystal x-ray and neutron refinements indicate that for as the as-grown crystals, whose properties are discussed in this manuscript, the vacancies are not completely ordered (see Table \ref{tab:refine}) in spite of slow cooling. Whether different heat treatments in the 150\,$^{\circ}$C to 400\,$^{\circ}$C temperature range can significantly change the degree of vacancy ordering requires further investigation. We note, however, that if the crystals are cooled too quickly from 1070$^{\circ}$C to room temperature, the two phase transitions at 140 and 100\,K are not visible in resistivity data.  Single crystal plates with typical dimensions of 5 $\times$ 5 $\times$ 2 mm$^3$ could be easily cleaved from the solidified boule. Because of the potential toxicity of Tl compounds, the samples were always handled with gloves, and often under a hood. The crystals, however, appeared to be quite stable in air and maintained a shiny surface even after exposure to ambient conditions for many hours. A powder x-ray diffraction (PXRD) slide was purposely left in air overnight, and there was no observable change in the diffraction pattern after this exposure. Energy dispersive x-ray spectroscopy (EDS) indicated that the crystals were Fe deficient with an approximate composition of TlFe$_{1.6 \pm 0.1}$Se$_2$ that was uniform throughout the crystal. Virtually the same iron site occupancy was also obtained from PXRD and single-crystal neutron and x-ray structural refinements (Fe$_{1.58}$) (See Table \ref{tab:refine}). For property measurements, crystals were selected using a microscope to avoid choosing samples with small amounts of iron metal attached. A bar magnet provided an initial check, followed by a magnetization curve at room temperature.  

\begin{table*}
\caption{Refined structural parameters for TlFe$_{1.6}$Se$_2$ from single-crystal x-ray diffraction data.  Space group $I4/m$ (No. 87), $a$ = 8.661(2)\AA, $c$ = 13.971(7) \AA, $T$ = 200\,K, R1  =  0.0341, wR$F^2$ = 0.0672, and $\chi^2$ = 3.63 for 755 $F_o > 4\sigma$($F_o$).}
\begin{tabular}[c]{|c|c|c|c|c|c|c|}
\hline
 & Wyckoff & $x$ & $y$ & $z$ & site  & $U_{eq}$ \\
atom &  position & & &  & occupation & (\AA$^2$) \\
\hline
Tl1	& 2$a$ &	0 &	0	& 0 &	1	& 0.0226(11)  \\
Tl2	& 8$h$ &	0.19492(18)	& 0.39585(18) &	0	& 1	&0.0204(6)   \\
Fe1	& 16$i$ &	0.0932(3) 	& 0.1970(3)	& 0.2478(2)	& 0.926(5) &  	0.0077(11)   \\
Fe2	& 4$d$	& 0 &	1/2 &	3/4	& 0.272(15) & 	0.081(13)   \\
Se1	& 4$e$	& 0	& 0	& 0.3645(2)   & 1 &	0.0117(14)  \\
Se2	& 16$i$	& 0.1992(2) &	0.3922(2) &	0.35537(9)  &	1 &	0.0107(8)   \\
\hline
\hline
atom &  $U_{11}$ &$U_{22}$ & $U_{33}$& $U_{12}$ & $U_{13}$ & $U_{23}$ \\
\hline
Tl1	& 0.0207(11)	& 0.0207(11)	&  0.0264(12)	&  0.00000	 & 0.00000	& 0.00000 \\
Tl2 &	0.0191(6)	  & 0.0198(6)	  &  0.0224(4)  &	-0.0006(6) &	0.00000	& 0.00000 \\
Fe1	& 0.0086(13)	& 0.0089(12)	&  0.0057(7)	& 0.0007(12) &	-0.0006(17) &	0.0001(12) \\
Se1	& 0.0136(13)	& 0.0136(13)	&  0.0080(17)	& 0.00000	   & 0.00000	& 0.00000 \\
Se2	& 0.0107(9)	  & 0.0117(9)	  &  0.0097(6)	& -0.0009(8) &	0.0019(9)	& -0.0009(8) \\
\hline
\end{tabular}
\label{tab:refine}
\end{table*}

\section{Experimental Details}

Resistivity, magnetization, specific heat and Hall measurements were made using commercial equipment from Quantum Design. A Physical Property Measuring System (PPMS) was used for the resistivity, specific heat and Hall measurements, and a SQUID magnetometer (MPMS) for magnetization and susceptibility measurements. For electrical transport measurements, 0.025\,mm diameter platinum wires were affixed to the crystals using Dupont 4929N silver paste. PXRD measurements were made in vacuum from room temperature to 11\,K using a PANalytical X'Pert Pro MPD equipped with a PHOENIX closed cycle cryostat using Cu $K_{\alpha}$  radiation. Above room temperature PXRD were collected with a similar instrument under flowing helium and the temperature was controlled by an Anton Paar XRK 900 furnace. Scanning electron microscope (SEM) and energy dispersive x-ray (EDS) measurements were made with a Hitachi TM-3000 tabletop microscope equipped with a Bruker Quantax 70 EDS system. Single crystal neutron diffraction measurements from 5\,K to 450\,K were made at HB-3A four-circle diffractometer at the High Flux Isotope Reactor at the Oak Ridge National Laboratory.  Superlattice structural model was also tested using single-crystal x-ray diffraction data at 200\,K collected with a Bruker SMART APEX CCD diffractometer with Mo-$K\alpha$ radiation.

\section{Results and Discussion}

\begin{figure}
	\centering
\includegraphics[width=3.2in]{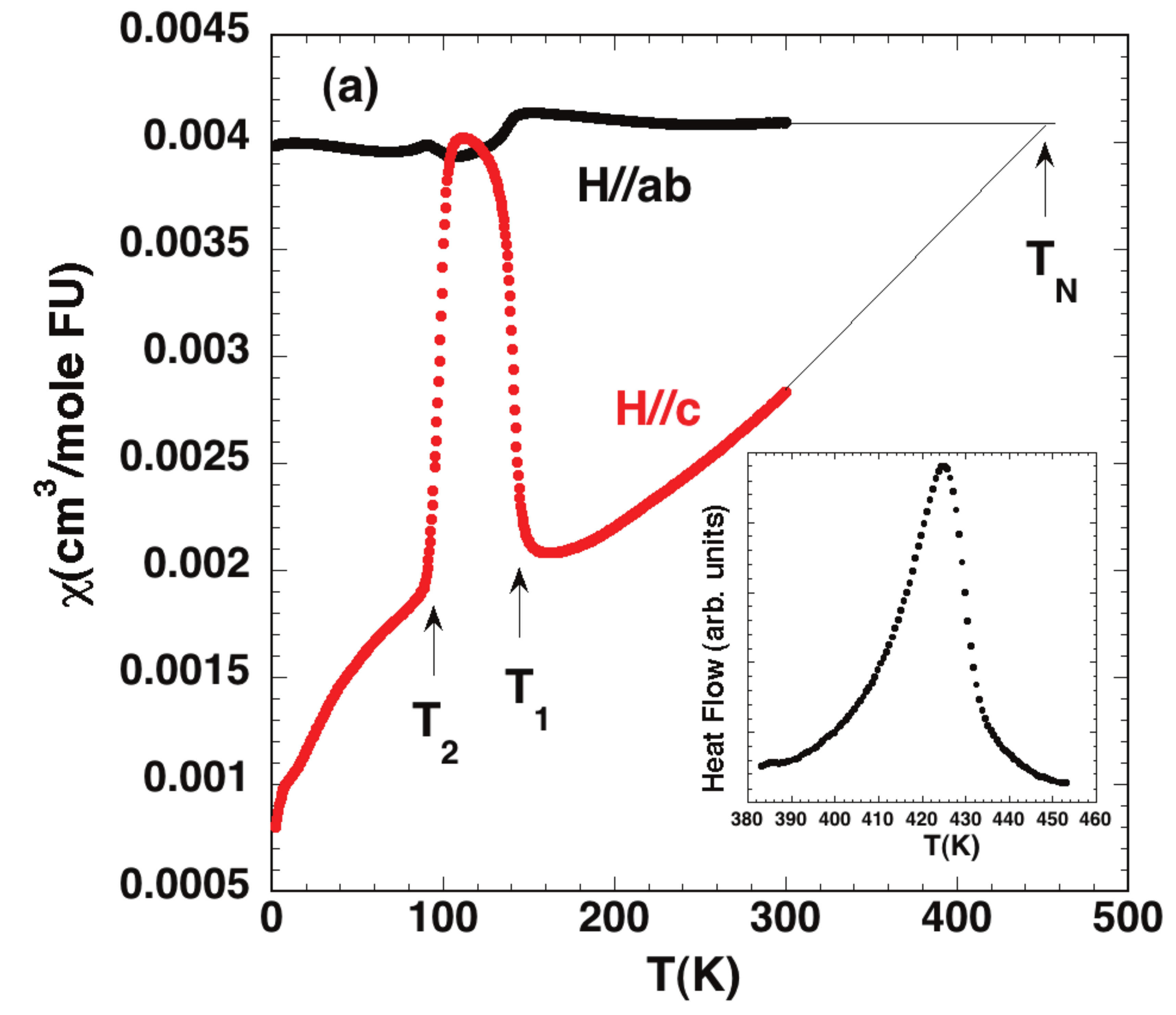}
\includegraphics[width=3.2in]{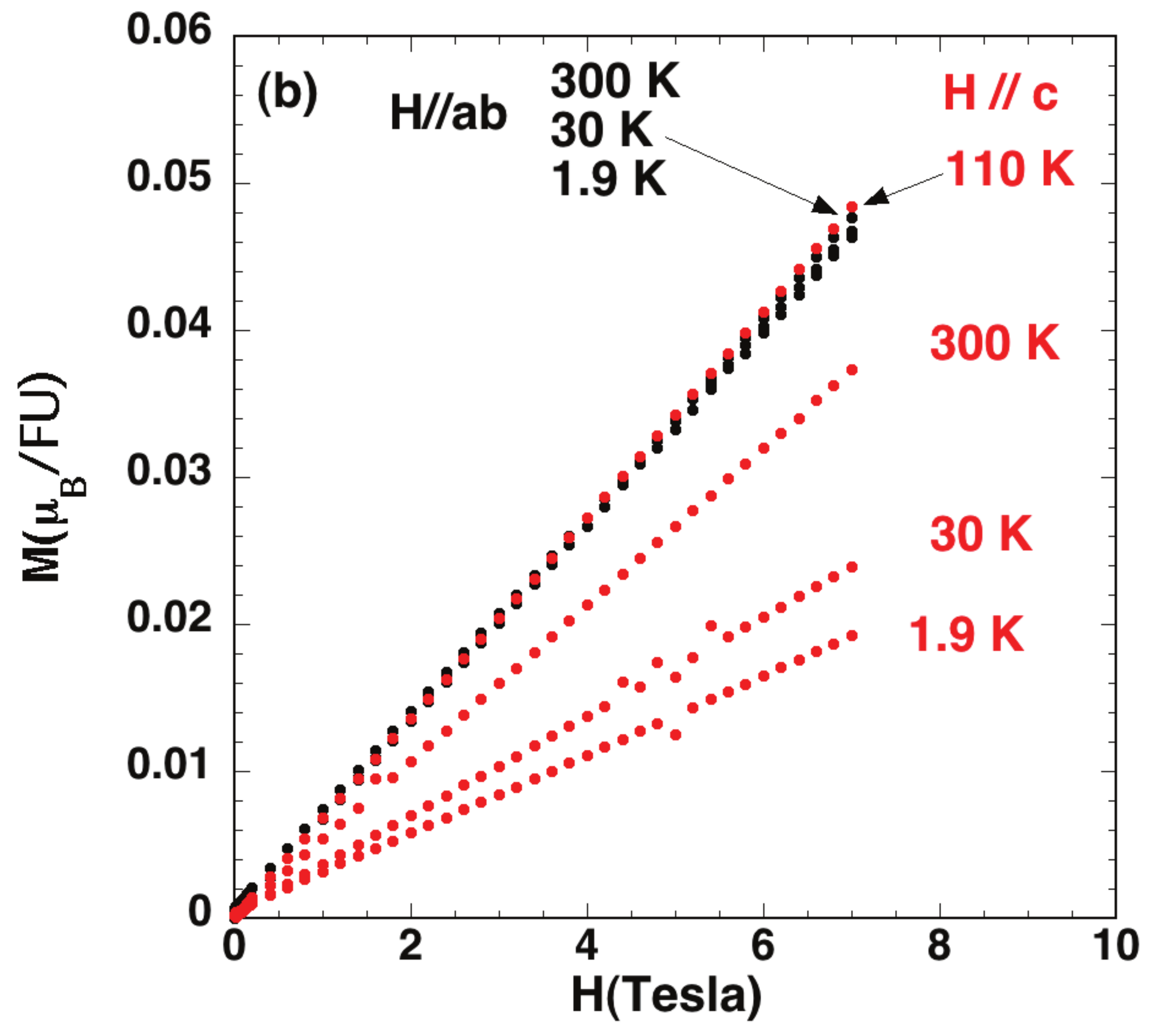}
\caption{(color online) (a) Magnetic susceptibility of TlFe$_{1.6}$Se$_2$ vs. temperature for \textbf{H} = 5T (ZFC). The Neel temperature, $T_N$, as well as two additional phase transitions at $T_1$ and $T_2$ are noted in the figure. The thermal signature near the Neel temperature obtained from a scanning calorimeter is shown in the inset. (b) Isothermal magnetization curves at temperatures above, between, and below the phase transitions at $T_1$ and $T_2$.}
	\label{fig:2}
\end{figure}

The magnetic susceptibility versus temperature is shown in Fig \ref{fig:2}a with an applied magnetic field, \textbf{H}, of 5 Tesla for \textbf{H}$\parallel$\textit{\textbf{c}} and \textbf{H} $\perp$  \textit{\textbf{c}}. Ignoring for the moment the two additional phase transitions, $T_1$ and $T_2$, the overall temperature dependence of the magnetic susceptibility in the two directions suggests antiferromagnetic ordering with the Fe spins aligned along the \textit{\textbf{c}} axis. If the high temperature susceptibility data is extrapolated to temperatures above 300\,K, the intersection of a linear extrapolation for each direction occurs near 450\,K. This is close to the Neel temperature, $T_N$ $\approx$ 430\,K for this compound as determined by neutron diffraction (see below) or by differential scanning calorimetry measurements (DSC) (see inset Fig \ref{fig:2}a). This conclusion is consistent with the earlier work of H\"{a}ggstr\"{o}m et al.,\cite{Haggstrom1986} who suggested antiferromagnetic order with spins along the \textit{\textbf{c}} axis based on single crystal Mossbauer data and the lack of attraction of the samples to a permanent magnet. Below room temperature, two additional phase transitions are observed at $T_1$ $\approx$ 140\,K and $T_2$ $\approx$100\,K for both \textbf{H}$\parallel$\textit{\textbf{c}} and \textbf{H} $\perp$  \textit{\textbf{c}} (see Fig. \ref{fig:2}a). To the best of our knowledge these two transitions have not been previously observed in this compound or related compositions with alkali metal substitution, and we are not aware of similar features in other itinerant antiferromagnets. The large and abrupt increase in $\chi$  at 140\,K followed by an abrupt decrease in $\chi$ near 100\,K is particularly odd. This feature was observed in several crystals with \textbf{H} $\parallel$ \textit{\textbf{c}} during both heating and cooling through this temperature range.  Isothermal magnetization data for both directions are shown in Fig \ref{fig:2}b for temperatures above, between, and below the two transitions. With \textbf{H} $\perp$ \textit{\textbf{c}} (\textbf{H} $\parallel$ \textit{\textbf{ab}}) the magnetization curves are linear and have almost identical slopes for all temperatures between 1.9 and 300\,K. For a simple antiferromagnet with spins along \textit{\textbf{c}}, an almost constant susceptibility below $T_N$ is expected in this direction.  Also, for \textbf{H} $\perp$ \textit{\textbf{c}} the magnetization curves all pass through the origin, which confirms that the crystals are free of contamination from iron metal. With \textbf{H}$\parallel$\textit{\textbf{c}} axis the magnetization curves are also linear from 300\,K down to $T$=1.90\,K. 

\begin{figure}
	\centering
\includegraphics[width=3.2in]{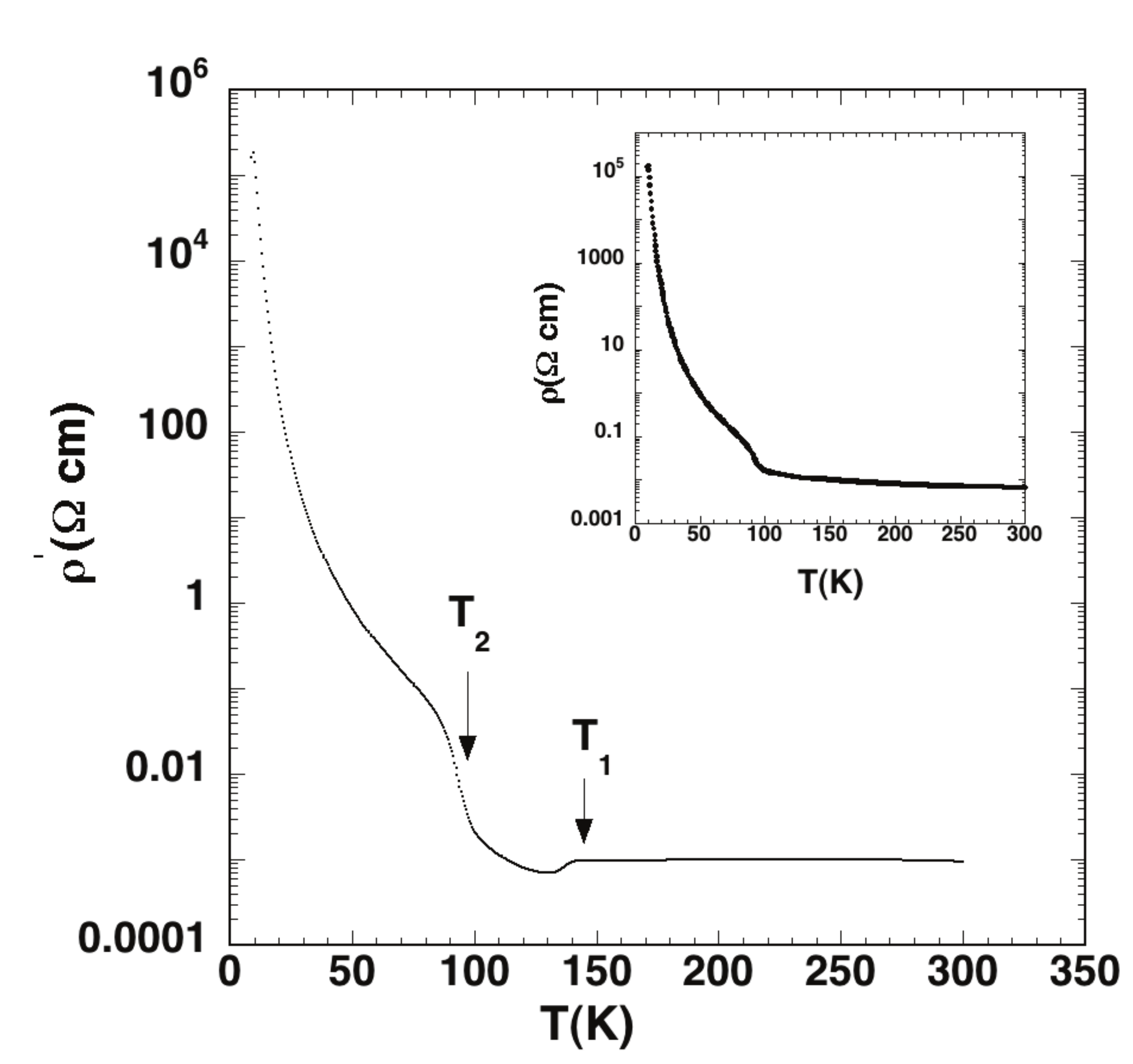}
\caption{Resistivity in the basal plane vs temperature for TlFe$_{1.6}$Se$_2$. The raw data is shown in the inset. The data between 150\,K and 300\,K is accurately described by simple activated behavior with an activation energy of 11\,meV. In the main panel, this activated behavior has been subtracted from all of the data so that the phase transition at $T_1$ is more easily seen.}
	\label{fig:3}
\end{figure}

In-plane resistivity data for TlFe$_{1.6}$Se$_2$ crystals are presented in Fig \ref{fig:3}. Above 150\,K, the resistivity data can be accurately described by a simple exponential of the form $\rho=\rho_0 e^{A/T}$ with $\rho_0$=0.00446\,$\Omega$-cm and $A$=124\,K, implying an activation energy for electron motion of 11\,meV. Just below $T_1$, the resistivity continues to increase (inset), but more slowly than expected from the high temperature behavior. This is illustrated in the main panel where the temperature dependence of the resistivity at high temperature is subtracted from all of the data. Near $T_2$ there is a rapid increase in the resistivity with decreasing temperature. At low temperatures, the resistivity continues to increase until it exceeds the limit of our instrumentation at 9\,K with a resistivity of 2$\times$10$^5\,\Omega$-cm.  The resistivity is not accurately described by simple activated behavior except for temperatures above 150\,K. The apparent electron concentration, $n$, as estimated from Hall data, varied from 5$\times$10$^{21}$ electrons/cm$^3$ at 150\,K to 2$\times$10$^{18}$ electrons/cm$^3$ at 50\,K.  The Hall data were converted to carrier concentration assuming a single carrier band and using a linear fit to the high field Hall data. Above 150\,K the Hall signal was too small to measure, probably indicating the presence of both electrons and holes; below 50\,K, the resistivity was too large. Although it is difficult to reliably interpret Hall data from multiband magnetic materials, the three order of magnitude drop in apparent carrier concentration between 150\,K and 50\,K is consistent with the three order of magnitude increase in the resistivity over this same temperature interval. 

\begin{figure}
	\centering
\includegraphics[width=3.2in]{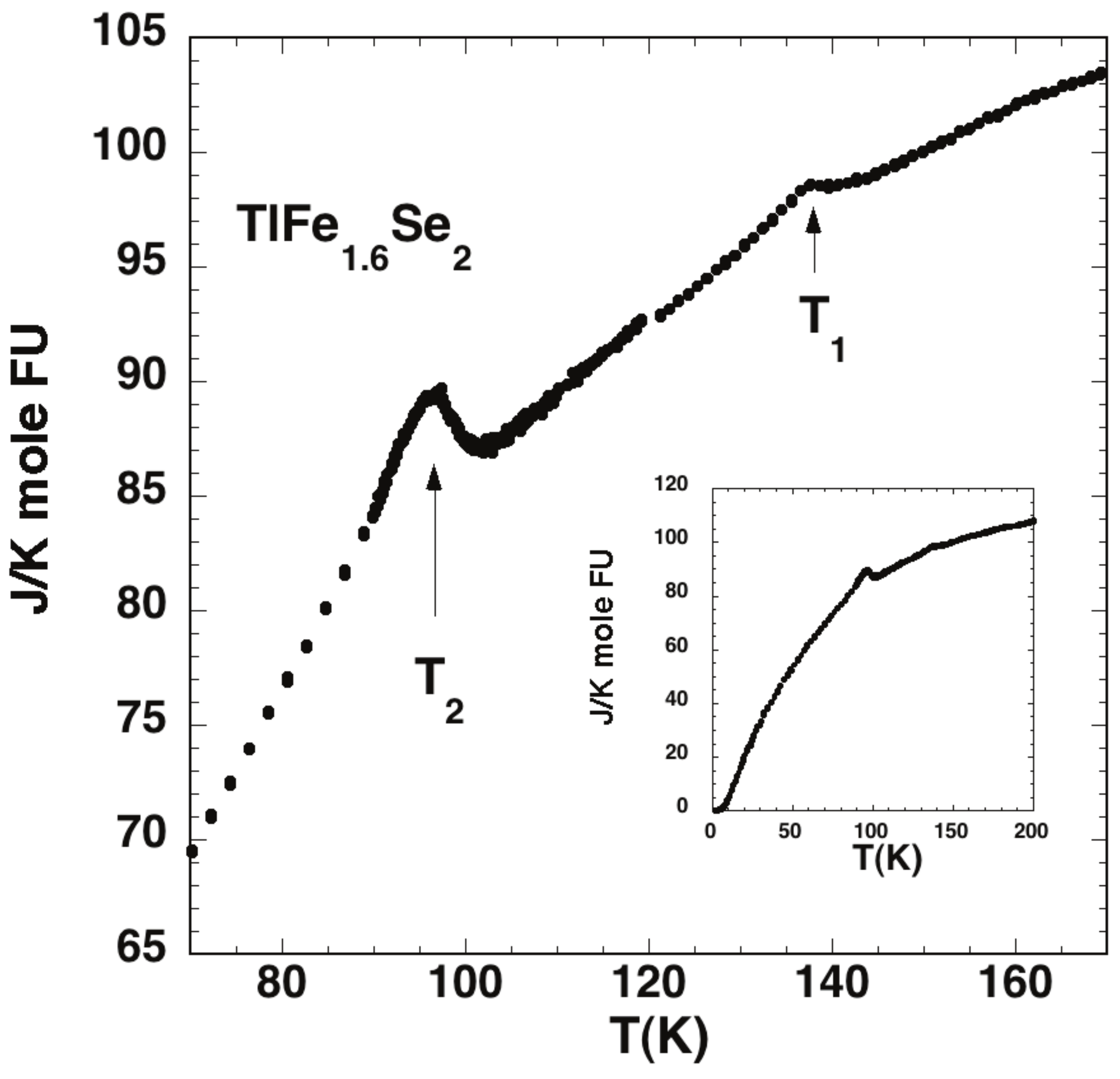}
\caption{Heat capacity versus temperature for TlFe$_{1.6}$Se$_2$. The inset shows all of the data from 200\,K to 2\,K. The main panel emphasizes the data near the two phase transitions $T_1$ and $T_2$.}
	\label{fig:4}
\end{figure}

Heat capacity ($C$) data for a TlFe$_{1.6}$Se$_2$ crystal from 200\,K to 2\,K are shown in the inset of Figure \ref{fig:4}. The main panel shows the heat capacity data in the vicinity of the two phase transitions $T_1$ and $T_2$. Clear peaks are observed at each transition. The entropy is estimated by plotting $C/T$ versus $T$ (not shown) and fitting a smooth polynomial through the data for temperatures just above and below each peak.  The area of each peak above this background is an estimate of the entropy associated with each phase transition.  The approximate change in entropy associated with each feature is:  $\Delta S$ = 0.06\,J/(K-mole-fu) for the $T_1$ transition and $\Delta S$ =0.54\,J/(K-mole-fu) for the $T_2$ transition. This entropy is small if compared to the entropy associated with complete magnetic order for a spin $\frac{1}{2}$ system Rln2= 5.76\,J/(K-mole spins). For comparison, the estimated entropy change associated with the DSC peak at $T_N$, shown in the inset of Figure \ref{fig:2}, is about 4\,J/(K-mole-fu).  Near this same temperature (430\,K), the superlattice peaks also disappear, so part of this entropy change is associated with a structural change (see below). As expected, analysis of the low temperature heat capacity data resulted in a value of $\gamma \approx$ 0 for the Sommerfeld coefficient, within our experimental error.

\begin{figure}
  \centering
     \subfigure{\includegraphics[width=1.67in]{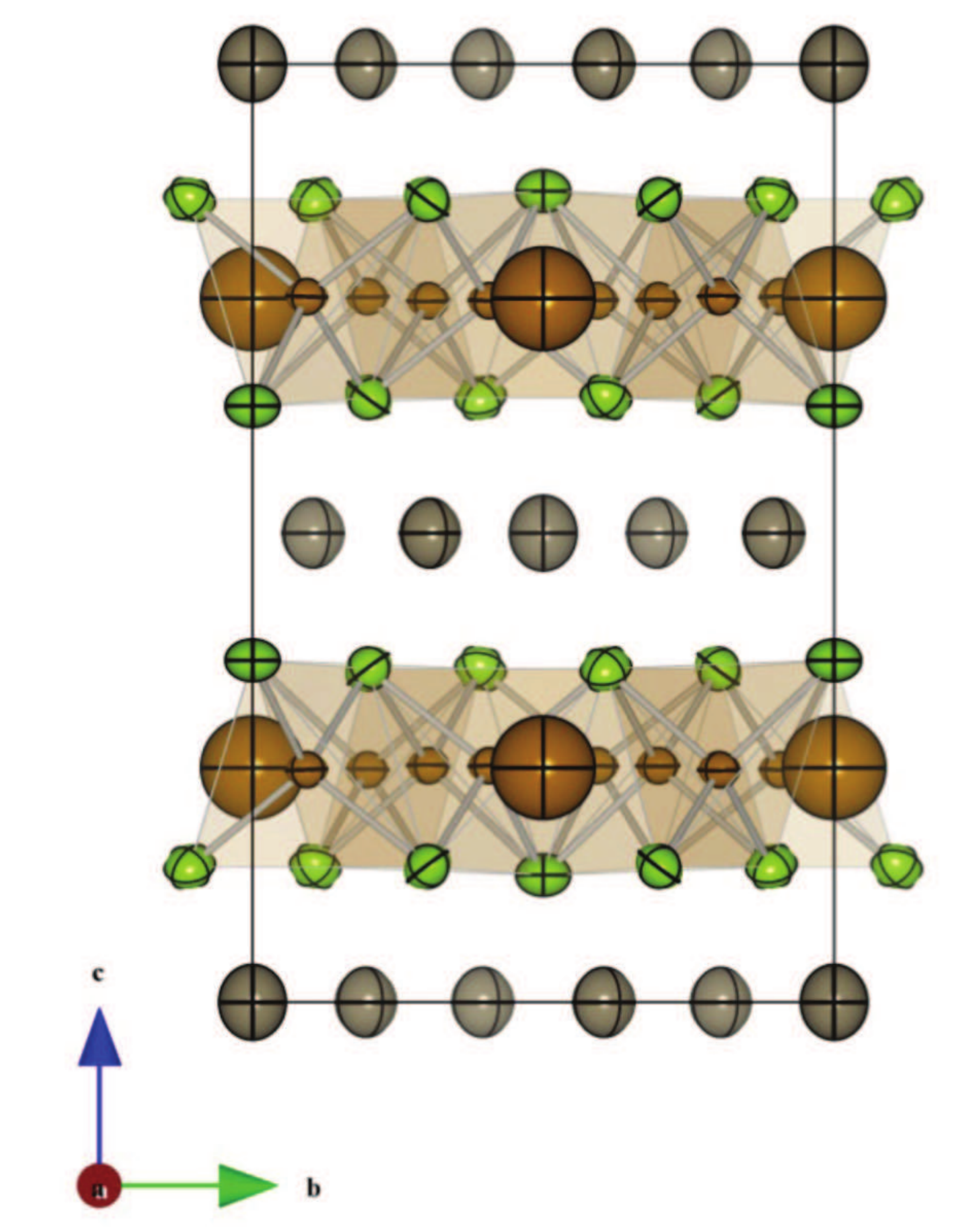}}
     \subfigure{\includegraphics[width=1.65in]{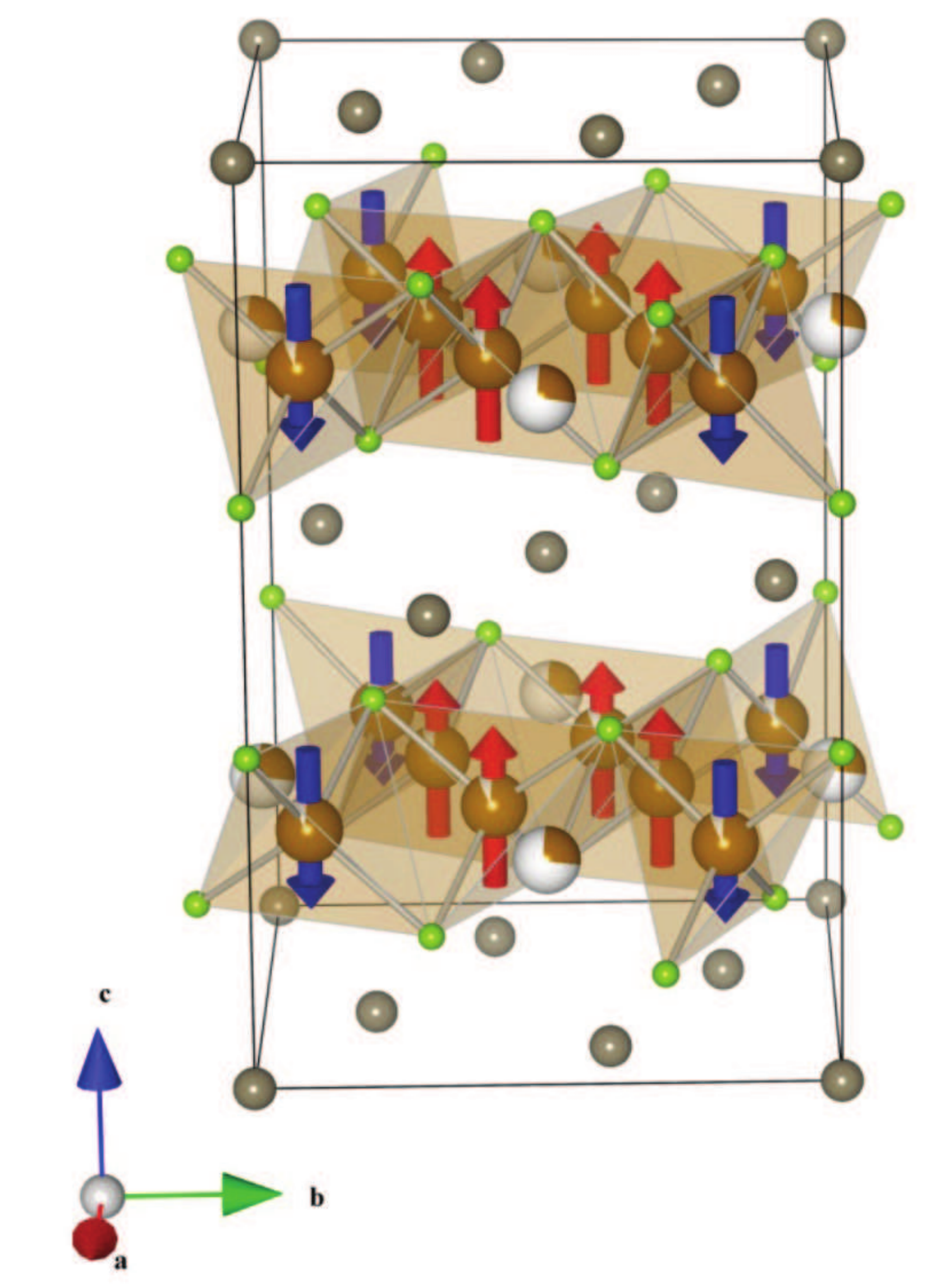}}
  \caption{(color online) Left panel: nuclear structure determined from single crystal X-ray diffraction, Tl (gray), Se (green), and Fe (brown). Atomic displacement parameters are drawn at 99\% probability.  Right panel: Tilted view of crystal structure with arrows providing a schematic representation of Fe spin orientation.  The amount of vacancies refined for the two iron sites are shown by the size of the white segment.}\label{fig:5}
\end{figure}

Single crystal neutron diffraction measurements were used to study both the magnetic structure and the crystal structure over the temperature range from 5 to 450\,K. At room temperature the crystal structure could be described using the tetragonal ThCr$_2$Si$_2$ structure with an additional superstructure to account for the partial ordering of Fe vacancies in the basal plane. This vacancy ordering produces additional weak superstructure reflections that are also observed in powder and single-crystal x-ray diffraction measurements (discussed below). This superstructure increases the \textit{\textbf{a}} lattice constant to $\sqrt{5}$\textit{\textbf{a}}, and hence results in a five fold increase in the cell volume, relative to the ThCr$_2$Si$_2$ subcell. The supercell (') is related to the subcell by \textit{\textbf{a}}' = 2\textit{\textbf{a}} + \textit{\textbf{b}}, \textit{\textbf{b}}' = -\textit{\textbf{a}} + 2\textit{\textbf{b}}, \textit{\textbf{c}}' = \textit{\textbf{c}}. The subcell is described with the space group symmetry $I4/mmm$ (No. 139) and the ordered vacancy superlattice structure is described with the space group $I4/m$ (No. 87). The strongest superlattice reflections are 0.7\% of the intensity of the strongest subcell peaks for single-crystal x-ray diffraction (Mo $K_{\alpha}$ radiation).  The nuclear structure as refined from the single-crystal x-ray data using the superlattice model is given in Table \ref{tab:refine}.  A principle difference between this structure and the ThCr$_2$Si$_2$ type structure, besides the obvious difference of the ordered vacancies, is a slight corrugation of the iron selenide tetrahedral layer, as can be seen in Figure \ref{fig:5}. The larger atomic displacement of Fe2 site implies more positional disordered than Fe1 site.  This gives rise to 5 unique Fe-Se bond lengths as compared to a single unique Fe-Se bond length in the ThCr$_2$Si$_2$ type structure.  The refined Fe content [1.590(3) Fe per formula unit] corresponds well to the Fe composition determined by EDS and powder x-ray diffraction. 

Comparing the x-ray and neutron diffraction data at room temperature, neutron data showed several Bragg peaks associated only with magnetic order. Using more than 50 superlattice peaks (contributed by both nuclear and magnetic scattering), the magnetic structure was solved at 300\,K, 140\,K, 115\,K, and 5\,K using the program Fullprof and symmetry group analysis.  Eight possible magnetic structures were used to refine the single crystal neutron data and the best fit is shown in Figure \ref{fig:5}. The magnetic peaks were included along with those from the vacancy ordered superlattice (about 240 in total) to refine both the magnetic and nuclear structures together as shown in Figure \ref{fig:5} using $I4/m$ space group symmetry. All the refinements have R-factors between 3.5-4.5\% ($\chi^2$ = 0.02-0.04). As expected from previous Mossbauer \cite{Haggstrom1986} and susceptibility data [Fig. \ref{fig:2}], the Fe spins point along the \textit{\textbf{c}}--axis and are antiferromagnetically aligned between layers. The single-crystal neutron refinements yield Fe vacancies on Fe1 ($\sim$10\%) and Fe2 ($\sim$70\%) positions similar to the x-ray refinement. Due to the low Fe2 occupation and the overlapping contribution of the nuclear superlattice peaks, the quality of our data did not allow refinement of any small possible moment contribution from the Fe2 site. Consequently, for the structure shown in Fig. \ref{fig:5}, no detectable magnetic moment occurs at the vacancy position (Fe2), and the magnetic structure is similar to that recently reported by Bao et al.\cite{Bao1102.0830} for the superconductor K$_{0.8}$Fe$_{1.6}$Se$_2$ and by others \cite{Ye1102.2882} for similar superconducting compositions.  The magnetic structure is a ``block checkerboard'' antiferromagnetic structure with 4 Fe spins near the center of the unit cell pointing down and 4 spins near the corners pointing up (see Fig. \ref{fig:5}). 

\begin{figure}
\includegraphics[width=3.2in]{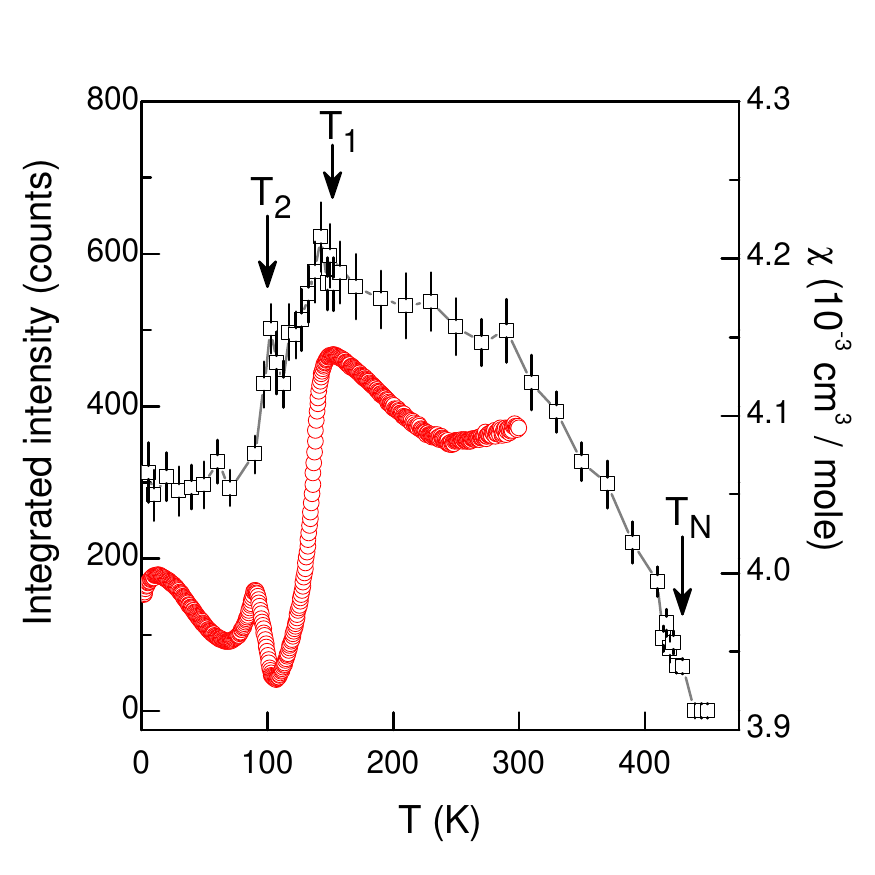}
\caption{(color online) Temperature dependence of (1 $\bar{2}$ 1) magnetic reflection in the $\sqrt{5}$\textit{\textbf{a}} $\times$  $\sqrt{5}$\textit{\textbf{a}} supercell notation (squares, left  axis), and magnetic susceptibility $\chi$ vs. temperature with \textbf{H} $\parallel$ \textit{\textbf{ab}} (circles, right axis, for \textbf{H} = 5\,T). Note the onset of magnetic reflection intensity near $T_N \approx$430\,K, and the peaks at both $T_1 \approx$ 140\,K and $T_2 \approx$100\,K. Also note the decreased intensity below 150\,K. This corresponds to a decrease in the magnitude of the Fe moments below 150\,K.}
	\label{fig:6}
\end{figure}

The same ``block checkerboard'' magnetic structure provides a good description of the available data from 300\,K to 5\,K. The moment magnitude at 300\,K is 1.72(6)$\mu_B$ per iron, which is consistent with the magnitude of the hyperfine field from Mossbauer data,\cite{Haggstrom1986} and is considerably less than value of 3.31$\mu_B$ found by Bao et al. for K$_{0.8}$Fe$_{1.6}$Se$_2$.\cite{Bao1102.0830}  All of the magnetic peaks found at 250\,K, including the (101) reflection, decrease in intensity below 140\,K, as illustrated in Fig. \ref{fig:6} for the (1$\bar{2}$1) reflection.  If the magnetic structure does not change, this corresponds to a decrease in the magnetic moment per Fe from a maximum of 2.07(9)\,$\mu_B$ at 140\,K to 1.31(8)\,$\mu_B$ at 5\,K.  For most ordered antiferromagnetic materials this intensity should normally saturate at about $T_N$/2. In the present case the intensity significantly decreases below 150\,K. Also apparent in Fig. \ref{fig:6} are small peaks in intensity near both $T_1$ and $T_2$. For comparison, the susceptibility data with \textbf{H}$\parallel$\textit{\textbf{ab}} (shown in Fig. \ref{fig:2}a) is expanded and replotted in Figure \ref{fig:6}. The correspondence between the magnetic order parameter and the susceptibility in this direction is striking.  The data in Fig. \ref{fig:6} either imply a decrease in the magnitude of each iron moment or a change in the magnetic structure that is not detectable with our available neutron data sets.  Further neutron experiments are planned.

\begin{figure}
\includegraphics[width=3.3in]{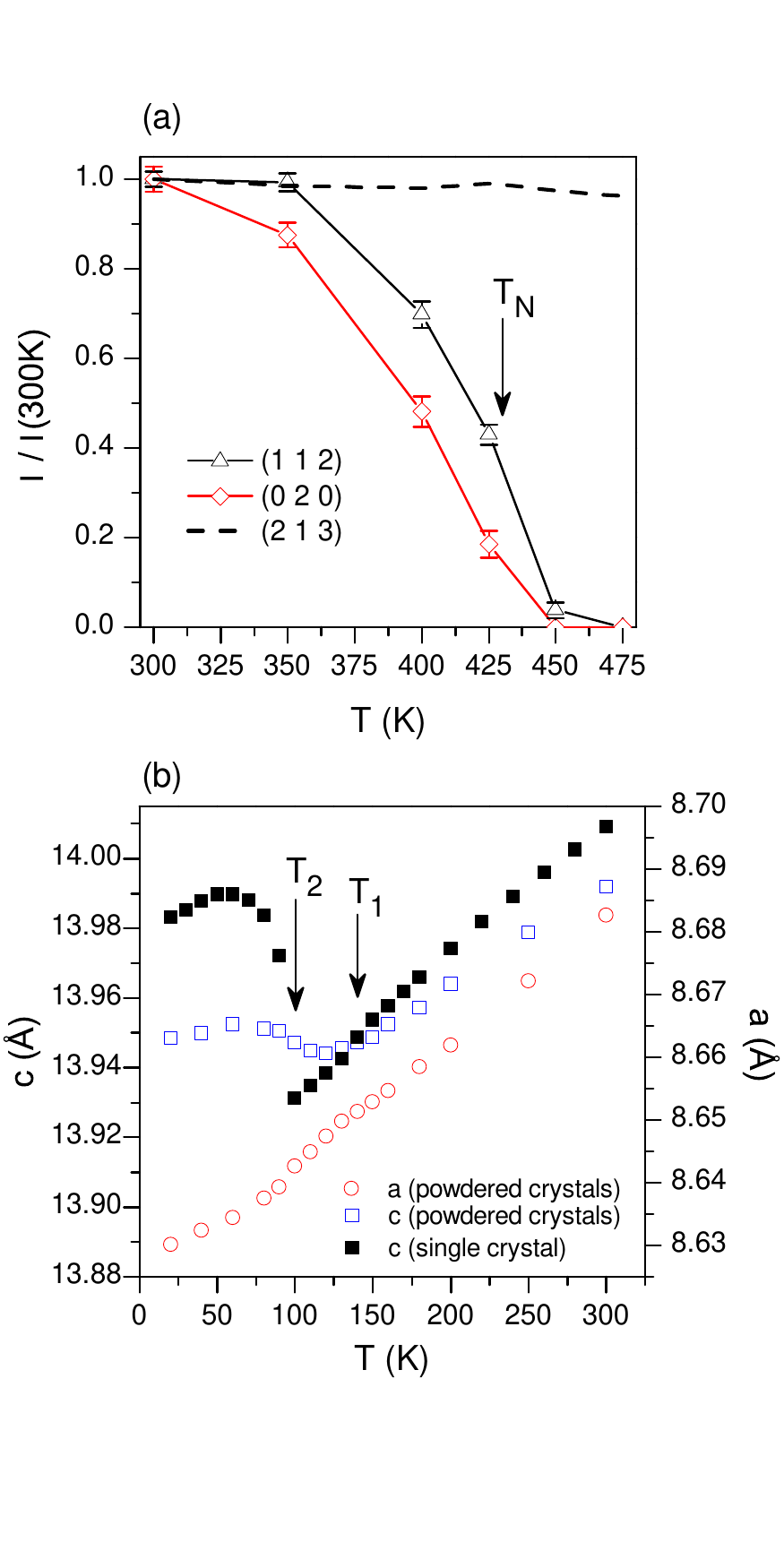}
\caption{(color online) (a) Normalized x-ray intensity of superlattice reflections (112) and (020) from TlFe$_{1.6}$Se$_2$ versus temperature. The supercell appears to disappear slightly above $T_N$. For comparison, only a small variation in intensity of the (213) reflection from the ThCr$_2$Si$_2$ subcell is observed (dashed line). (b) Variation of the \textit{\textbf{c}} and \textit{\textbf{a}} lattice constants with temperature. The approximate temperatures for the two phase transitions $T_1$ and $T_2$ are also shown.}
	\label{fig:7}
\end{figure}

The crystal structure of TlFe$_{1.6}$Se$_2$ as a function of temperature was studied using both single-crystal x-ray and neutron diffraction and PXRD, to relate changes in crystal structure to changes in the magnetism and other properties. At temperatures near or slightly above $T_N$, the superlattice reflections disappear, as is illustrated in Figure \ref{fig:7}. These results are similar to what happens in the superconductor K$_{0.8}$Fe$_{1.6}$Se$_2$.\cite{Bao1102.0830} For TlFe$_{1.6}$Se$_2$, the vacancies are disordered above 450\,K, and an average ThCr$_2$Si$_2$ structure is recovered. From room temperature to 11\,K, the crystal structure is described by the  $\sqrt{5}$\textit{\textbf{a}} $\times$ $\sqrt{5}$\textit{\textbf{a}} supercell of the ThCr$_2$Si$_2$ structure (Fig. \ref{fig:1}). Figure \ref{fig:7}b shows the temperature dependence of the \textit{\textbf{c}} lattice constant determined by measuring several (0 0 l) reflections from the surface of a single crystal platelet using a powder x-ray diffractometer. An abrupt and dramatic change in \textit{\textbf{c}} is observed at $T_2$, and a more subtle anomaly occurs at $T_1$. The correspondence of these features and the magnetic phase transitions demonstrate the strong interplay between the crystal structure and magnetism in this material.  Also shown in Figure \ref{fig:7}b are lattice parameters \textit{\textbf{a}} and \textit{\textbf{c}} determined from full-pattern refinements of PXRD data from crystals that had been ground into fine powder. The resulting powder was severely strained (as evidenced by broadened lineshapes); however, anomalous behavior is observed in both lattice parameters in the temperature range around $T_1$ and $T_2$. The decrease in \textit{\textbf{a}} and increase in \textit{\textbf{c}} near 100\,K result in a smooth temperature dependence for the unit cell volume (not shown). \textit{\textbf{c}}-axis measurements on single crystals near $T_2$ show some thermal hysteresis ($\sim$10\,K), and two lattice constants are observed at some temperatures around $T_2$. This suggests that this transition is likely first order.

\section{Summary and Conclusions}

The structural, magnetic, thermal and transport properties of single crystals of TlFe$_{1.6}$Se$_2$ are reported. The compound orders antiferromatically with $T_N$ $\approx$ 430\,K as determined from single crystal neutron diffraction. At or slightly above this same temperature, superlattice reflections develop on cooling, which correspond to the partial ordering of vacancies in the Fe layer. Below $T_N$ the crystal structure can be described as a $\sqrt{5}$\textit{\textbf{a}} $\times$ $\sqrt{5}$\textit{\textbf{a}} supercell of the ThCr$_2$Si$_2$ structure, as first shown by H\"{a}ggstr\"{o}m.\cite{Haggstrom1986} From single crystal x-ray and neutron diffraction, the ``vacant'' Fe site is 30 \% Fe-occupied and the ``filled'' Fe site is 90\% Fe-occupied in TlFe$_{1.6}$Se$_2$.  The antiferromagnetic structure is commensurate with the lattice with iron spins pointing along the \textit{\textbf{c}} axis. The detailed structure is sketched in Figure \ref{fig:5} and it is similar to the magnetic structure recently reported by Bao\cite{Bao1102.0830} for the K$_{0.8}$Fe$_{1.6}$Se$_2$ superconductor. TlFe$_{1.6}$Se$_2$, however, is an insulator at low temperatures. Upon cooling the TlFe$_{1.6}$Se$_2$ crystals below room temperature two new phase transitions are observed at $T_1$ $\approx$ 140\,K and $T_2$ $\approx$ 100\,K. The two transitions are evident in resistivity, heat capacity, and magnetic susceptibility measurements, and in the magnetic order parameter determined from neutron scattering. There is no detectable change in crystal symmetry at $T_1$ and $T_2$, only a jump in the \textit{\textbf{c}} lattice parameter at $T_2$, and unusual curvature in both lattice constants in the 100-150\,K temperature region. There is also no detectable change (from neutron scattering) in the overall arrangement of the moments in the magnetic structure, only a change in the magnitude of the magnetic moment per iron. At 300\,K, the refined Fe moment is 1.72(6)$\mu_B$, at 140\,K it is 2.07(9)$\mu_B$, at 115\,K the moment is 1.90(9)$\mu_B$, and at 5\,K it is 1.31(8)$\mu_B$. Near room temperature this compound is a decent electrical conductor with a resistivity of 0.006\,$\Omega$-cm that increases only slightly with temperature down to 150\,K. It is likely that the Fe d electrons strongly participate in a Fermi surface in this temperature region. Below 100\,K, the resistivity rapidly increases with temperature with the loss or localization of carriers and eventually the Fermi surface is destroyed. From this perspective, it is perhaps not too surprising that the moment per iron changes as the temperature is lowered. Exactly how this happens is not understood. Large magnetoelastic coupling, small changes in the Fe-Se distance, and the localization of carriers are likely responsible for the striking properties below 150\,K, but further research is needed before a detailed microscopic explanation can be proposed.

\section{Acknowledgements}

It is a pleasure to acknowledge useful discussions with Mark Lumsden and David Singh, and the technical assistance of Larry Walker and Andrew Payzant, and the ORNL glass shop.  Research was supported in part by the Division of Materials Sciences and Engineering, Office of Science, U.S. Department of Energy (BCS, MAM, AFM, AS).  The research at ORNL's High Flux Isotope Reactor was sponsored by the Scientific User Facilities Division, Office of Basic Energy Sciences, U. S. Department of Energy.

\end{document}